\documentclass{article}
\usepackage{listings}
\usepackage{graphicx}
\usepackage{epsfig}
\usepackage{inconsolata}
\usepackage{float}
\usepackage{authblk}
\begin{document}

\title{A highly scalable approach to solving linear systems using two-stage multisplitting}
\author[1]{Nick Brown\footnote{Corresponding author: +44 (0) 131 650 6420, nick.brown@ed.ac.uk}}
\author[1]{J. Mark Bull}
\author[1]{Iain Bethune}
\affil[1]{Edinburgh Parallel Computing Centre, James Clerk Maxwell Building, Kings Buildings, Edinburgh}
\date{}
\maketitle

\begin{abstract}
{\it Iterative methods for solving large sparse systems of linear equations are widely used in many HPC applications. Extreme scaling of these methods can be difficult, however, since global communication to form dot products is typically required at every iteration.

To try to overcome this limitation we propose a hybrid approach, where the matrix is partitioned into blocks. Within each block, we use a highly optimised (parallel) conventional solver, but we then couple the blocks together using block Jacobi or some other multisplitting technique that can be implemented in either a synchronous or an asynchronous fashion. This allows us to limit the block size to the point where the conventional iterative methods no longer scale, and to avoid global communication (and possibly synchronisation) across all processes. 

Our block framework has been built to use PETSc, a popular scientific suite for solving sparse linear systems, as the synchronous intra-block solver, and we demonstrate results on up to 32768 cores of a Cray XE6 system. At this scale, the conventional solvers are still more efficient, though trends suggest that the hybrid approach may be beneficial at higher core counts.}
\end{abstract}


\section{Introduction}

As systems approach Exaflop performance, the core counts which are typically used to solve problems will increase dramatically. To exploit Exascale resources effectively, existing algorithms and implementations must be re-examined to break the tight coupling that often exists between parallel processes. We concentrate on algorithms for the solution of linear equations \emph{Ax = b}, where \emph{A} is a large, sparse $n \times n$ matrix and both \emph{x} and \emph{b} are vectors. Solving such large, sparse linear systems is critical in many fiends of scientific computing and so solution methods which can efficiently run at Exascale are crucial.

In Bethune et al.\cite{asyncpaper} we have shown that for iterative solution methods, instead of relying on synchronous communication for exchanging updated vector information one can gain an improvement in performance and resilience at large core counts by using asynchronous communication where processes need not wait for their neighbours at each iteration and the calculation is therefore not globally synchronised. However, our experiments only considered a point Jacobi iterative solution method which typically converges very slowly.

In this paper we extend our work by applying Jacobi's algorithm, and other multisplitting techniques, at the block level using asynchronous communication. Within each block we use much more computationally efficient Krylov Subspace (KS) methods in an attempt to combine the advantages of scalability and resilience from the asynchronous methods with the computational efficiency of KS methods. We investigate the performance and scalability of these different approaches on up to 32768 cores on a Cray XE6, and consider the effects on performance of tuning various parameters in the hybrid algorithm.

The rest of the paper is organised as follows: Section 2 reviews the existing algorithms, and in Section 3 we present our hybrid algorithm. Section 4 reports experimental results, and Section 5 draws some conclusions and considers future work.  

\section{Background} 

\subsection{Jacobi's algorithm}

Jacobi's algorithm is the simplest iterative solution method. Whilst the convergence rate of this algorithm is inferior to other, more complex methods, the fact that the only global communication required is in computing the residual makes it highly scalable. This aspect of the algorithm makes it of interest to Exascale where problems must be decomposed over very many cores.

For a linear system, $Ax=b$, one starts with a trial solution $x_{0}$ and generates new solutions, iteratively, according to $x_{i}^{(k)} = \frac{1}{a_{ii}}(b_{i}-\sum_{i != j}a_{ij}x_{j}^{(k-1)})$ where $k$ is the iteration number. One stops iterating when the norm of the global residual is smaller than a specific tolerance, $\parallel Ax-b \parallel_{2} < tol $. 

If $A$ represents a discretised local operator, resulting from the application of a small stencil over a domain, then the parallel implementation is very simple. In this paper we solve the 3D Laplace equation, $\bigtriangledown^{2}u=0$ using a seven point stencil, and the Jacobi algorithm can be written in pseudo code as:
\begin{lstlisting}[basicstyle=\ttfamily\small,breaklines=true]
for all grid points
	unew(i,j,k) = 1/6 * ( u(i+1,j,k)+u(i-1,j,k)+
			u(i,j+1,k)+u(i,j-1,k)+
			u(i,j,k+1)+u(i,j,k-1) )
\end{lstlisting}

The calculation at each point only requires data from the nearest neighbouring points in each dimension. When solving in parallel using a regular domain decomposition, a halo swap is performed between neighbouring processes once per iteration to communicate the values required for the next iteration. Whilst this halo swap communication involves, at most, the number of neighbouring processes (six in the example above) it can still be a performance bottleneck when done synchronously as each process must wait for all of their neighbours to communicate before continuing with the next iteration. Alternatively, the swap can be done asynchronously, where processes will not wait for communication to complete and instead will use whatever halo values that they have for the next iteration, which may not be the most up-to-date ones as messages from the neighbouring processes might still be in flight. Choosing the communication method is a trade-off between communication time and data accuracy; one can wait longer for more recent data or complete communications much faster but potentially have older data to operate with. It is important to note that, whilst the asynchronous halo swap case is a different mathematical algorithm, it will still converge under appropriate conditions\cite{convergetheory}.

In our previous work studying the point Jacobi algorithm using asynchronous communication \cite{asyncpaper}, we found that for smaller numbers of cores (up to 4096) it is generally more important to have the latest data available and synchronous communication performs favourably. However, as one increases the number of cores, the asynchronous communication choice becomes more attractive. At these larger core counts, although the age of the data means more iterations are needed overall the reduced communication overhead more than makes up for this.

\subsection{Krylov subspace methods}

Krylov subspace methods are more efficient algorithms for solving sparse linear systems. A large number of iterative algorithms exist, based upon Krylov subspaces, with common ones being Conjugate Gradient (CG)\cite{cg} and Generalized Minimal RESidual method (GMRES)\cite{gmres}. It is very common for scientific codes to use one of these solver kernels and some excellent toolkits exist to facilitate this. One such library; the Portable, Extensible Toolkit for Scientific Computation (PETSc)\cite{petsc} is a suite of data structures and routines developed by Argonne National Laboratory for use in finding parallel solutions to scientific problems modeled by partial differential equations. Toolkits such as PETSc support a variety of Krylov subspace methods and allow for scientific programmers to utilise these more advanced, performant, iterative solvers. Additionally, low level aspects of parallelism such as process synchronisation and communication are abstracted away by the library.

 \begin{center} 
\begin{figure}[htb]
\begin{center}\includegraphics[scale=0.8]{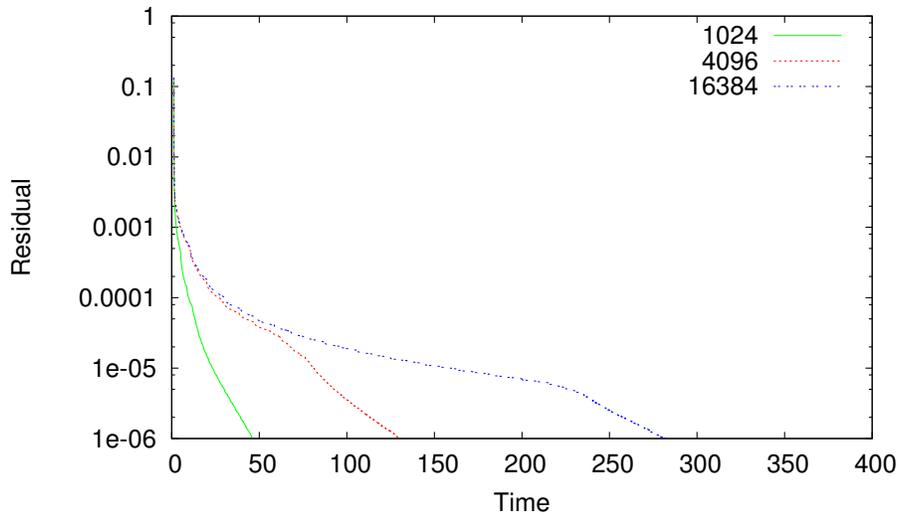}\end{center}
 \caption{GMRES Relative residual against time for different core counts}
    \label{fig:perf-pure}
\end{figure}
\end{center}

Figure \ref{fig:perf-pure} shows the relative residual against time when using GMRES from PETSc for solving the 3D Laplace's equation with Dirichlet boundary conditions on a Cray XE6 with different numbers of cores, using one MPI process per core. Since in each case the local domain is $50 \times 50 \times 50$ points per core, this is a weak scaling test. It can be seen that as the core count increases the performance drops. When run over 4096 cores it takes 2.88 times as long to solve the system than it did over 1024 cores (although the system is larger due to weak scaling.) This performance decrease gets worse, on 16384 cores the runtime is 6.34 times that of 1024 cores to find a solution. Some, but not all of the increase in execution time is due to a rise in the number of iterations required. The average time per iteration over 1024 cores is 0.05 seconds, when run on 16384 cores this has increased to 0.1 seconds. This suggests that existing, conventional, iterative solution methods will not scale to the core counts that will be likely required at Exascale. We were unable to run this test on 32768 cores as the code failed with an out of memory error.

\begin{table}[H]
\centering
\begin{tabular}{ | c | c | }
\hline
Cores \quad&\quad Iterations to solution\\
\hline			
1024 \quad&\quad  772\\
4096 \quad&\quad  1404\\ 
16384 \quad&\quad 1767\\ 
\hline
\end{tabular}
\caption{GMRES iterations to relative residual of $10^{-6}$}
\label{tbl:gmits}
\end{table}

Table \ref{tbl:gmits} illustrates the number of iterations required to reach a relative residual of $10^{-6}$ for this problem when solved using GMRES. It can be seen that, as one increases the number of cores then the problem is becoming more difficult to solve as more iterations are needed. The reason for this is that, as we weak scale and the problem size becomes larger, the spectral radius of the matrix also increases which requires more iterations to convergence.

\subsection {Multisplitting algorithms}
Often a matrix \emph{A} has a natural splitting into $A=M - N$ where linear systems involving \emph{M} can be solved easily. A splitting of \emph{A} in this manner results in a linear system of the form $Mx^{k+1} = Nx^{k} + b$ which starts with an initial guess $x^{0}$ and the system converges if the spectral radius of the iteration matrix ($M^{-1}N$) is less than 1. A number of common splittings of the matrix \emph{A} have been developed and many of these have been studied in \cite{multisplit}. If a matrix can be split in multiple ways, i.e. $A = M_{i} - N_{i}$, for $i= 1, \ldots, k$. then the multisplitting of \emph{A} is defined by  $B= \sum_{i=1}^{k}D_{i}M_{i}^{-1}N_{i}$. This is an attractive approach because the work for each individual splitting can be assigned to distinct processes.

Two stage algorithms further the matrices $M_{i}$ into two further components and perform a number of inner iterations on these. A variety of work has been done investigating factors such as whether to fix the number of inner iterations, or to vary it as the number of outer iterations progresses \cite{multi-innerits}. Another choice is that of communication and \cite{multi-comm} considers making the the outer iterations asynchronous, where processors are allowed to start the computation of the next inner iteration without waiting for the completion of the same outer iteration on other processes. In this case, the previous iteration's data is not guaranteed to be available to all processes:  the latest received, but not necessarily most up to date, values will be used. 

Two stage multisplitting such as \cite{multi-simple} concentrates on using a simple iterative method such as Jacobi to solve the inner stage. However the performance of these approaches is severely limited and so a two stage Krylov multisplitting algorithm \cite{multi-ks} has been developed. Note the much of the literature in this area is of a theoretical nature, typically focusing of proving convergence criteria, and does not often address how the splittings, communication modes, or the number of inner iterations affect real world performance or scaling. 

\section{Hybrid algorithm}

In \cite{asyncpaper} we showed that as we increased the core count, using asynchronous communication in the point Jacobi algorithm becomes beneficial. However, point Jacobi is of little practical interest: the benefits gained from using asynchronous communication are far outweighed by the slow convergence of the algorithm. 

\begin{center} 
\begin{figure}[H]
\begin{center}\includegraphics[scale=0.5]{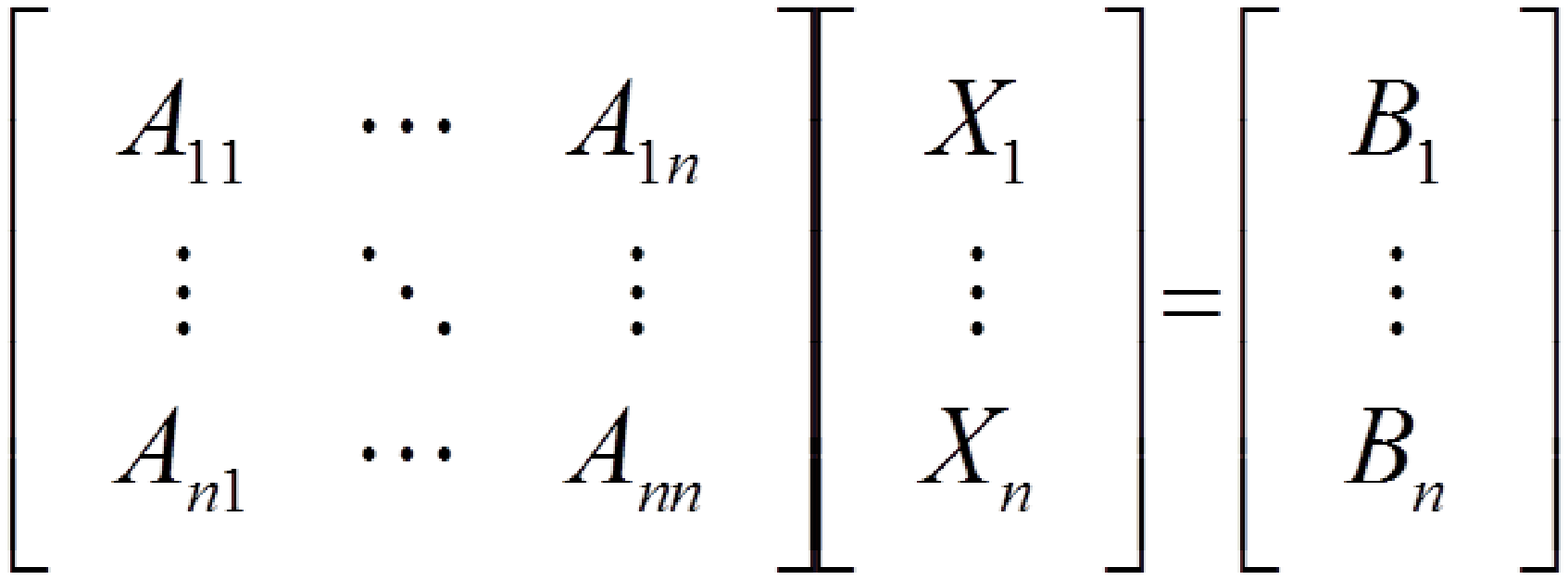}\end{center}
 \caption{Linear system rewritten as blocks}
    \label{fig:blocksystem}
\end{figure}
\end{center}

It is possible to write a linear system in block form, where each block $X_{i}$ is made up of a number of individual elements, $x_{i}$ and the matrix $A$ is split as shown in figure \ref{fig:blocksystem}. The Jacobi iterative algorithm can then be rewritten in terms of blocks as $X_{i}^{(k)} = A_{ii}^{-1}(B_{i}-\sum_{j!=i}A_{ij}X_{j}^{(k-1)})$.

For each block we now need to solve the linear system $A_{ii}X_{i}^{(k)} = (B_{i}-\sum_{j!=i}A_{ij}X_{j}^{(k-1)})$. One can see that if the rows of $A$ are distributed across processors in the same way as $x$, then the only communication required to assemble the right hand side is collecting the necessary blocks $X$. Assuming domain decomposition and local operators, this requires only halo swaps between neighbouring domains with the inner solve of the block systems being fully local.

Our hybrid approach is a two stage multisplitting algorithm where the inner solve of each block can be done by a group of processes using some efficient algorithm such as CG or GMRES. The inter- and intra-block concerns are separated and whilst the intra-block solve is done synchronously, the halo swap between block groups after each block Jacobi iteration can be done asynchronously. Listing \ref{lst:pseudoblock} illustrates a pseudo code example of this block scheme. This inter-block communication is implemented in our framework on a processor to processor basis rather than funnelling all communications via a master, and uses the same techniques as in \cite{asyncpaper}. If new data is available after an asynchronous halo swap then this will be used in the next inner block solve, and if not, then the inner solve will continue using existing data from a previous iteration. The asynchronous halo swap maintains \emph{R} buffers (\emph{R}=100 in the experiments for this paper) which hold sent and received halo data for each neighbour along with the MPI asynchronous communication requests. At all stages there are \emph{R} outstanding receives registered for each neighbour and at each halo swap old sends are cleaned up and their buffer space returned to the pool. If there is buffer space available then an additional neighbour asynchronous halo send is issued and a check is then made as to how many pending receives have completed since the last halo swap. If any have completed then data from the latest of these receives is copied into the actual solution halo and pending receives are then reissued.   

The relative residual is obtained for each block and from this the global relative residual is the square root of the sum of the square of block local relative residues. Communication of the block local residues can be done using a tree-based asynchronous reduction, as described in \cite{asyncpaper}, which means that processes have an estimate of the global residual, but it might be some number of iterations old. It should be noted that the inbuilt MPI synchronous allreduce outperforms our own asynchronous allreduce and this is because MPI collective communications are heavily tuned and often optimised in hardware. The new MPI 3.0 standard contains non-blocking collective communications which could be used instead of our asynchronous reduction, and would likely be more efficient 

\begin{lstlisting}[caption=Pseudo code of the block scheme, label=lst:pseudoblock, captionpos=b, basicstyle=\small,breaklines=true]
loop while global residual is larger than threshold
	perform inner KS block solve
	halo swap at block level (asynchronous communication)
	recompute global residual (asynchronous communication)
\end{lstlisting}

As discussed in Section 2, point Jacobi scales well, but is slow to converge. More performant methods such as CG and GMRES as seen in Section 3, might perform well on medium core counts but do not scale to the level required for Exascale. Our proposed asynchronous two stage multisplitting algorithm aims to achieve the best of both worlds; the scalability of Jacobi (at the outer stage) and the good convergence of KS methods (at the inner stage.)

\subsection{The block solver framework}

We have built a framework, which uses PETSc for the inner solver and allows us to control many parameters to investigate these two stage multisplitting algorithms. Using PETSc we can easily select between different inner solvers which scale reasonably well.

Whilst the framework itself supports a variety of pluggable problem definitions, as discussed in this paper we concentrate on the 3D Laplace equation. A variety of parameters must be selected before the run: the number and dimensions of groups, number of inner iterations and size of the block overlap. In our framework, communication involving overlapping solution values is combined with that of normal block halo swapping. Therefore, asynchronous halo swapping also results in asynchronous resolution of overlapping solution values. 

\subsection{Choice of parameters}

\begin{center} 
\begin{figure}[H]
\begin{center}\includegraphics[scale=0.8]{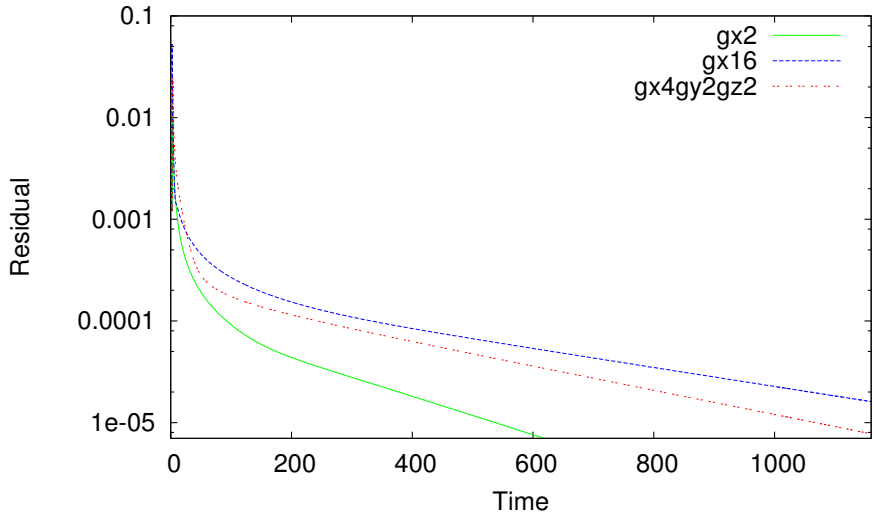}\end{center}
 \caption{Relative residual against time for different group configurations, inner iterations=10 and no overlap}
    \label{fig:perf-groups}
\end{figure}
\end{center}

By separating the concerns of the inter- and intra-block levels we can configure each independently for optimised performance, scalability and resilience. The first, and most obvious, choice is the size of the inner blocks and how the global domain is decomposed. Figure \ref{fig:perf-groups} illustrates the relative residual against time for different block configurations over 4096 cores with a problem size of $50 \times 50 \times 50$ per core running on a Cray XE6 system. In this figure three configurations are plotted, 2 blocks in the x-dimension (gx2), 16 blocks in the x-dimension (gx16) and 16 blocks distributed more evenly between the x-, y- and z-dimensions (gx4gy2gz2.) It can be seen that the optimal number of blocks is two and distributing blocks over all three dimensions is preferable to all blocks in one single dimension. The fact that the optimal number of blocks is two with this core count and problem size supports the intuition that when the limit of conventional solvers are reached, the solution space can then be partitioned into blocks for increased scalability and at this test configuration the scalability limit of the inner solver has not yet been reached. Solving to a relative residual of $10^{-5}$, gx=2 takes 1106 outer iterations, gx16 requires 2984 outer iterations and gx4gy2gz2 results in 2298 outer iterations. It can be seen that, as we increase the number of groups then this increases the number of outer iterations required to solve the problem. This supports the intuition that, as one splits the matrix for these groups the spectral radius of the iteration matrix increases, and hence more iterations are required to reach convergence. It is an interesting result that distributing the blocks over the dimensions (gx4gy2gz2) rather than in one dimension (gx16) is preferable - both in terms of runtime and number of outer iterations. This is likely due to the fact that distribution in one dimension only means that at each halo swap blocks receive data from most two neighbours and progress made in the right most block takes a number of iterations to reach the left most block. Contrasted against distributing the blocks amongst all dimensions, halo swaps involve more neighbours and-so updates can percolate throughout the system much faster.

\begin{center} 
\begin{figure}[H]
\begin{center}\includegraphics[scale=0.8]{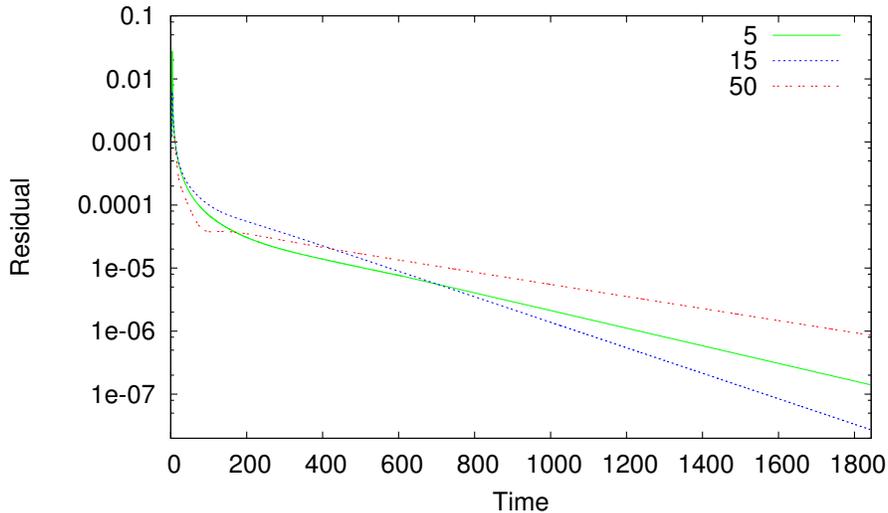}\end{center}
 \caption{Relative residual against time for numbers of max inner iterations, gx=2,gy=gz=1 and no overlap}
    \label{fig:perf-itrate}
\end{figure}
\end{center}

Another important inter-block factor is deciding what the inner solve threshold should be, i.e. when the inner solve is terminated and a halo swap performed before the next block Jacobi iteration. This is a trade-off between the frequency of halo swapping between blocks and the progress made at each inner solve. Figure \ref{fig:perf-itrate} illustrates the relative residual against time for different numbers of maximum inner GMRES iterations for a problem size of $50 \times 50 \times 50$  per core running on 4096 cores. It can be seen that the best choice depends heavily on the progression of the solution with time. Both 5 and 50 maximum inner iterations perform favourably initially, but when the relative residual reaches a certain point then performance decreases considerably, and the choice of 15 maximum inner iterations is better for this problem when solving to higher precision. The optimal choice for the inner block solver convergence criterion is problem dependent and currently we allow for the number of inner iterations to be fixed with one specific value for the entire solve, and this value must be tuned manually. The number of outer iterations is heavily linked to the number of inner iterations; with a smaller number of inner iterations then the number of outer iterations is large and a larger number of inner iterations results in a smaller number of outer iterations. For instance, solving to a relative residual of $7 \times 10^{-6}$ results in 3000 outer iterations when the number if inner iterations is set to 5, 1035 outer iterations with 15 inner iterations and 360 outer iterations with 50 inner iterations.  

\begin{center} 
\begin{figure}[H]
\begin{center}\includegraphics[scale=0.8]{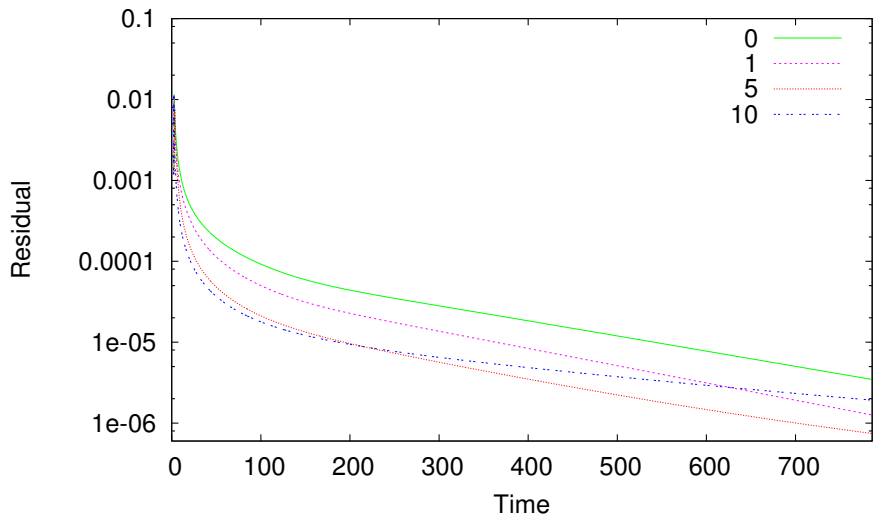}\end{center}
 \caption{Relative residual against time when overlapping solution space, gx=2, gy=gz=1 and inner iterations=10}
    \label{fig:perf-overlap}
\end{figure}
\end{center}

At the inter-block level, the ability to overlap the solution space at the block level can help accelerate block convergence. Multisplitting was first introduced in \cite{multisplit} where it was shown that overlapping blocks are still guaranteed to converge, and may do so faster. A scheme is needed for determining the overall value of an element by weighting contributions from multiple overlapping blocks and when considering a complex data decomposition, several groups can overlap the same element. Figure \ref{fig:perf-overlap} illustrates the relative residual against time for different overlaps over 4096 cores. It can be seen that some overlapping of the solution space does improve performance, although the optimal choice is difficult to determine a priori. The performance when overlapping the solution spaces by ten elements, which initially is very good, starts to degrade as we go to more accurate solutions. Generally it has been found that overlapping by between one and five elements in each dimension is a reasonable choice. The amount of overlap has an impact upon the number of outer iterations required to solve the problem. Solving to a relative residual of $2 \times 10^{-6}$ , with no overlap we see 2341 outer iterations, with an overlap of 1 element this decreases to 1771 outer iterations, an overlap of 5 elements to 1330 outer iterations and 10 elements to 1920. This helps explain the results in Figure \ref{fig:perf-overlap}; it can be seen that block convergence accelerates both in terms of runtime and outer iterations as the width of the overlap is increased to a point, however after reaching this point then additional overlap starts to slow block convergence. This can be explained by the fact that, regardless of the overlap, the solution size remains the same, so by increasing the amount of overlap increases the number of unknowns on each process. The number of additional unknowns per process in 3D, assuming neighbours on all sides, is given by o(2xy+2yz+2zx), where o is the size of the overlap, and x, y and z are the number of unknowns in each dimension. In our experiments x,y,z = 50 (125000 unknowns per process) Therefore with an overlap of 1, there are 15000 additional unknowns per process, an overlap of 5 results in 75000 additional unknowns per process and overlap of 10 requires 150000 additional unknowns. 

\section{Results}

We evaluated the block Jacobi approach for the Laplace equation on a Cray XE6. For each core count we have evaluated a conventional GMRES solve against block Jacobi using synchronous and asynchronous halo swapping. There are two blocks in the x-dimension and the local problem size is $50 \times 50 \times 50$ per core using weak scaling. All other parameters have been tuned to be the most optimal settings for this problem at that core count.

\begin{center} 
\begin{figure}[H]
\begin{center}\includegraphics[scale=1]{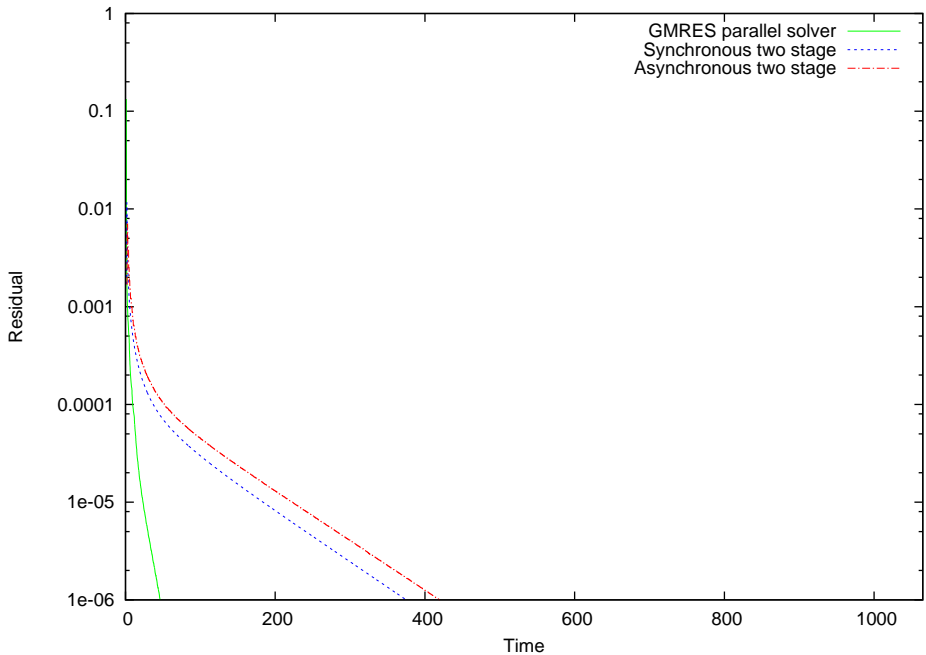}\end{center}
 \caption{Relative residual against time for 1024 cores}
    \label{fig:perf-1024}
\end{figure}
\end{center}

Figure \ref{fig:perf-1024} illustrates the performance results over 1024 cores. It can be seen that at this core count solving the entire system conventionally outperforms the block Jacobi approach quite by some margin. This is no real surprise because, as seen in Figure \ref{fig:perf-pure}, GMRES is still scaling well at this core count. It can also be seen that asynchronous block Jacobi communication is slightly slower than synchronous communication with the synchronous version requiring 1395 outer iterations until convergence ($10^{-6}$) and the asynchronous version 1567 outer iterations. Due to the small number of cores, the cost of data being out of date in the asynchronous version and hence requiring more outer iterations to solution outweighs the benefits gained by not having to wait for communications to complete in a block halo swap. 

\begin{center} 
\begin{figure}[H]
\includegraphics[scale=1]{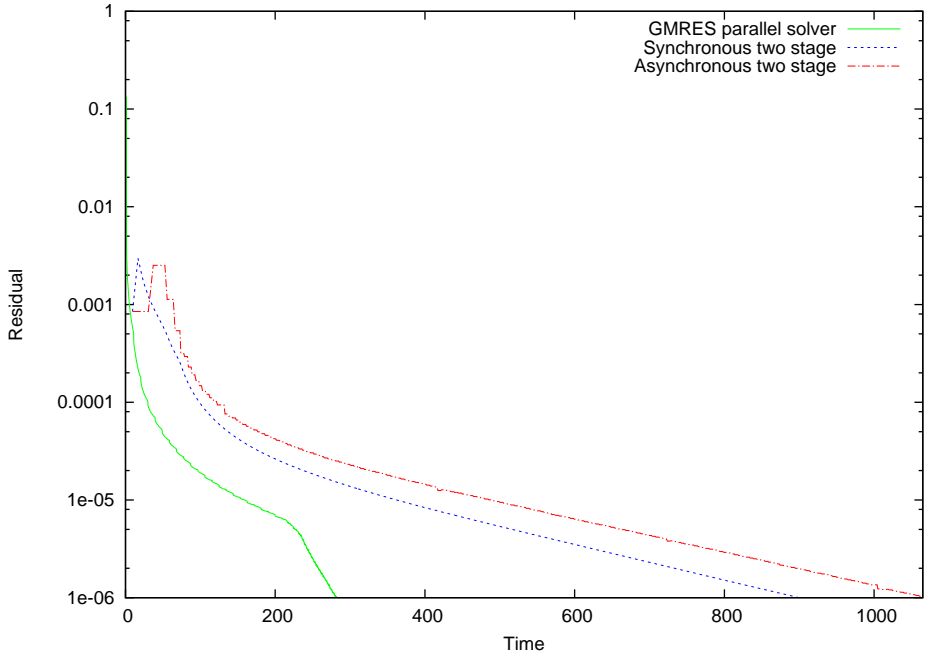}
 \caption{Relative residual against time for 16384 cores}
    \label{fig:perf-16384}
\end{figure}
\end{center}

The performance results for 16384 cores are illustrated in Figure \ref{fig:perf-16384}. One can see that the performance difference between solving the entire system conventionally using a single KS method and splitting it into blocks is smaller than on 1024 cores. The synchronous block version requires 680 outer iterations to solution and the asynchronous version 810 outer iterations to convergence. It should be noted that the actual number of iterations is smaller for the 16384 run compared to the 1024 run - this is simply because we have tuned the run parameters to be most efficient at the specific scale and for this larger run we are using 25 inner iterations compared to 10 which results in fewer outer iterations. Nevertheless, the results at this core count indicate that the cost of using out of date data and the increased number of outer iterations still outweighs the benefits gained from asynchronous communications. 

Table \ref{tbl:ratio} lists the ratio of execution time for each version compared to its run time on 1024 cores. It can be seen that the versions employing the block Jacobi approach scale better with the numbers of cores than the conventional solver version. Table \ref{tbl:itrate} illustrates the iteration rate (number of inner iterations times number of outer iterations) for each version. It can be seen that not only does the block Jacobi algorithm, regardless of communication method, sustain a higher iteration rate but, comparing the iteration rate between 1024 and 16384 cores we can see that when using only GMRES the rate it decreases by 47.5\%, whereas for Block Jacobi using synchronous communuication it only drops by 25.6\% and for asynchronous 26.2\%. 

\begin{table}[H]
\centering
\begin{tabular}{ | c | c | c | c | }
\hline
Cores \quad&\quad Conventional \quad&\quad Group Sync \quad&\quad Group Async\\
\hline			
1024 \quad&\quad  1 \quad&\quad 1 \quad&\quad 1\\
4096 \quad&\quad  2.88 \quad&\quad 1.47 \quad&\quad 1.57\\ 
16384 \quad&\quad 6.24 \quad&\quad 2.39 \quad&\quad 2.55\\ 
\hline
\end{tabular}
\caption{Ratio of execution time to 1024 cores for each version}
\label{tbl:ratio}
\end{table}

\begin{table}[H]
\centering
\begin{tabular}{ | c | c | c | c | }
\hline
Cores \quad&\quad Conventional \quad&\quad Group Sync \quad&\quad Group Async\\
\hline			
1024 \quad&\quad  19.20 \quad&\quad 26.11 \quad&\quad 26.21\\
4096 \quad&\quad  16.57 \quad&\quad 20.72 \quad&\quad 20.78\\ 
16384 \quad&\quad 10.05 \quad&\quad 19.41 \quad&\quad 19.34\\ 
\hline
\end{tabular}
\caption{Iteration rate (iterations per second)}
\label{tbl:itrate}
\end{table}

\subsection{Higher core counts}
\begin{center} 
\begin{figure}[H]
\begin{center}\includegraphics[scale=1]{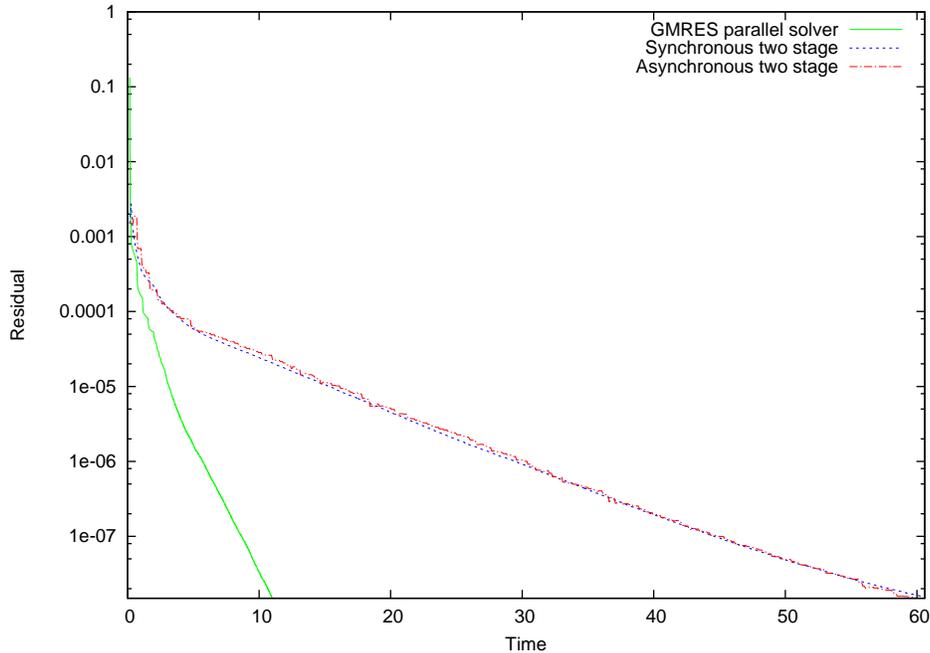}\end{center}
 \caption{Relative residual against time for 32768 cores}
    \label{fig:perf-32768}
\end{figure}
\end{center}

It has been mentioned that a conventional solve using GMRES would not run over 32768 cores with a local problem size of $50 \times 50 \times 50$ elements due to memory constraints, so we have run the same problem with a local problem size of $20 \times 20 \times 20$ using 32768 cores. The synchronous version required 350 outer iterations and the asynchronous version 392. It can be seen from Figure \ref{fig:perf-32768} that whilst over this much smaller problem size the version using only GMRES performs much better, it is interesting to see that the performance of synchronous and asynchronous block Jacobi is much more comparable. This result is not surprising as it is often seen with large numbers of cores asynchronous algorithms scale better then their synchronous counterparts. With greater levels of parallelism, more cores are waiting for each other in the synchronous halo swap and factors such as network latency can become an issue. As explained in Section 3, each block level halo swap involves each process on a face communicating with its neighbouring process on the face of the corresponding block, rather than funneling all data through a master process for each block which would not be scalable. This means that synchronous block level halo swaps may only proceed when all of these individual processes have completed communications. Instead, when each process on the block face communicates in an asynchronous fashion, the halo data used is the latest to be received but potentially not the most up to date in the system. Nevertheless, even though the data might not be entirely up to date, this becomes more favourable compared to synchronously waiting for all block level halo communications to complete. It is our expectation that experiments carried on larger numbers of cores will start to see the asynchronous block level communication out perform the synchronous communications.   

\section{Conclusions}

We have implemented a framework for solving large, sparse linear systems via a block Jacobi method with asynchronous or synchronous communication. From our investigations the most important tunable parameter has been the method of communication and we have seen that this has an impact upon the overall performance. For smaller core counts using GMRES without block Jacobi outperformed our approach quite considerably, however as the number of cores has been increased this performance gap has closed because the block Jacobi approach scales better.  With core counts up to and including 16384, synchronous block Jacobi communication outperforms asynchronous communications. However, as we move to 32768 cores, albeit with a smaller local problem size, both perform equally well.

Additional advantages of asynchronous block Jacobi at Exascale include resilience to performance faults such as slow cores, slow nodes or slow network links. Traditional synchronous solution methods are limited to the iteration rate of the slowest process but those based on asynchronous communication can continue to progress even if a group of processors are not contributing for some period of time.     

From our results we have seen that as we increase the core count both the block Jacobi method and asynchronous communication are starting to become competitive. It is our intent to further investigate with much larger core counts and, as traditional solution methods reach their limits, partition the solution space into larger numbers of blocks to enable further scaling. We will investigate the performance for more realistic problems, and try to reduce the number and sensitivity of the tunable parameters.  Another possibility is to investigate auto-tuning of these parameters at runtime. Not only will this avoid the difficulty of estimating these values a priori, it will also allow for the chosen parameters to be modified during the solve: as we saw in Figure \ref{fig:perf-itrate} certain choices perform well initially but then worsen as the solution progresses. Selecting appropriate values at different stages of the calculation should result in an improvement in overall performance.

\end{document}